\documentclass{Interspeech2024}
\pdfoutput=1
\interspeechcameraready 
\usepackage{microtype}
\usepackage{graphicx}
\usepackage{acronym}
\usepackage{subfigure}
\usepackage{booktabs}
\usepackage{hyperref}

\usepackage{colortbl}
\usepackage{soul}
\usepackage[dvipsnames]{xcolor}
\usepackage{color, xcolor} 
\usepackage[utf8]{inputenc}
\usepackage{algorithm}
\usepackage{algpseudocode}

\usepackage{amsmath}
\usepackage{amssymb}
\usepackage{mathtools}
\usepackage{amsthm}
\usepackage[capitalize,noabbrev]{cleveref}

\theoremstyle{plain}

\theoremstyle{definition}

\theoremstyle{remark}

\usepackage[textsize=tiny]{todonotes}

\title{Pre-training Feature Guided Diffusion Model for Speech Enhancement}

\name[]{Yiyuan}{Yang}
\name[]{Niki}{Trigoni}
\name[]{Andrew}{Markham}

\address{
  Department of Computer Science \\ University of Oxford, UK}
\email{\{yiyuan.yang,niki.trigoni,andrew.markham\}@cs.ox.ac.uk}


\begin{document}

\maketitle

\begin{abstract}
Speech enhancement significantly improves the clarity and intelligibility of speech in noisy environments, improving communication and listening experiences. In this paper, we introduce a novel pretraining feature-guided diffusion model tailored for efficient speech enhancement, addressing the limitations of existing discriminative and generative models. By integrating spectral features into a variational autoencoder (VAE) and leveraging pre-trained features for guidance during the reverse process, coupled with the utilization of the deterministic discrete integration method (DDIM) to streamline sampling steps, our model improves efficiency and speech enhancement quality. Demonstrating state-of-the-art results on two public datasets with different SNRs, our model outshines other baselines in efficiency and robustness. The proposed method not only optimizes performance but also enhances practical deployment capabilities, without increasing computational demands.
\end{abstract}

\section{Introduction}

In real-world application scenarios and hardware devices, clean acoustic sources are inevitably contaminated by environmental noise, speaker interference, and codec degradation~\cite{gay2012acoustic,lemercier2023storm}. This pollution leads to lower signal-to-noise ratios, which in turn degrade the quality of signal collection and adversely affect subsequent recognition or monitoring tasks. Consequently, denoising and enhancing the collected speech information becomes a critical necessity. Speech enhancement technology is designed to improve speech clarity and quality by separating the human voice from various types of noise signals, thus improving listening comfort. Various automatic speech enhancement techniques are currently in widespread use~\cite{saleem2019review}.

Facing to this issue, various classical algorithms were developed, aiming to leverage the statistical differences between clean speech signals and environmental noise~\cite{gerkmann2011unbiased, gerkmann2018spectral}. However, complexity is further compounded by factors such as the variability of noise sources, the wide spectrum of noise types, and the particular difficulties presented by non-stationary superimposed noise~\cite{saleem2019review}. On the other hand, recently proposed machine learning-based algorithms extract more effective latent features by learning acoustic representations and noise from data. Specifically, the methodologies can be categorized into two different types: discriminative and generative approaches~\cite{tai2023revisiting}.

Discriminative models employ supervised learning-based machine learning algorithms to map noisy speech to its corresponding clean speech target, and training with labeled speech samples~\cite{wang2018supervised}. These methods are capable of extracting effective features from a diverse set of clean and noisy speech pairs in various domains, e.g. time, spectrum, cepstrum, or spatial distribution~\cite{gerkmann2018spectral}. Discriminative models, while effective in specific situations, often lack understanding of other real data. This limitation means that it is difficult to go beyond their training data and adapt to new noise types, reverberation, or signal-to-noise ratio (SNR)~\cite{welker2022speech}. It is also a concern to introduce the risk of unexpected speech distortion, which may negate the benefits of noise reduction. Moreover, reliance on a large number of labeled datasets for training makes deployment more complex and prevents them from being widely used in real applications.

On the other hand, recent advances in generative models have significantly bridged these gaps. Generative speech enhancement algorithms use generative models (e.g., VAE~\cite{fang2021variational}, GAN~\cite{routray2022phase}, flow-based models~\cite{nugraha2020flow}, and diffusion model~\cite{tai2024dose}) to explicitly or implicitly learn the distribution of clean speech. Unlike discriminative models that learn a mapping from noisy speech to clean speech, generative models treat the speech enhancement task as a generative problem. Compared to discriminative ones, generative models demonstrate greater robustness to unseen scenarios and hold the potential to produce more natural speech due to optimizing prediction performance while simultaneously learning the distribution of inputs. They also exhibit superior noise handling and generalization capabilities. Generative models learn the overall distribution of speech data during training. Therefore, they can capture and replicate subtle details and features of speech, thus preserving more of the original sound quality~\cite{welker2022speech}.

In particular, diffusion model-based speech enhancement methods have recently received more attention, demonstrating their ability to achieve improvements in acoustic synthesis~\cite{lemercier2023storm,lu2021study,richter2023speech,welker2022speech-arXiv}. Compared to other generative models, those based on diffusion models exhibit superior enhancement effects, demonstrate stable training, and possess strong generalization capabilities regarding model architecture~\cite{tai2023revisiting,tai2024dose}. Specifically, approaches that utilize the diffusion process can be divided into two categories. The first category uses a diffusion-based vocoder to synthesize clean speech from an unconditional prior, with a conditioning network processing noisy input to denoising and improving speech for the vocoder~\cite{serra2022universal}. The second category directly models background noise or reverberation during the forward diffusion process and generates clean speech in the reverse process. This method can be implemented either in the time domain or the frequency domain~\cite{welker2022speech}.

Despite having been well improved and theoretically supported, diffusion-based approaches are still not the first choice for generative speech enhancement algorithms in real applications. Performance, efficiency, training steps, and inference time have impeded their performance in speech enhancement~\cite{tai2023revisiting}. In detail, \emph{\textbf{how to improve efficiency}} is the first challenge. To ensure high sampling quality and performance, diffusion model-based methods typically use hundreds of training/inference steps. In contrast, if inference is accelerated simply only by reducing the number of samples, the effectiveness of noise reduction decreases exponentially~\cite{tai2023dose}. Secondly, \emph{\textbf{how to choose complementary guidances}} to further improve the enhancement performance. Researchers have discovered that conditional diffusion models can achieve more stable and superior enhancement effects. However, they sometimes lose the semantic correspondence between the conditional noisy speech and the synthesized clean speech. Therefore, choosing the appropriate condition/guidance to steer the generative process is crucial~\cite{tai2023revisiting}.

This paper introduces a pretraining \underline{\textbf{f}}eature-guided \underline{\textbf{u}}nified diffusion model for efficient \underline{\textbf{s}}peech \underline{\textbf{e}}nhancement, named \textbf{FUSE}. Utilizing two complementary acoustic features as outlined~\cite{yang2023ssl}, we incorporate spectral features into a VAE, leveraging its latent feature map for the conditional diffusion model to improve the efficiency. Additionally, we utilize another set of learning-based pre-trained features as guidance/conditions during the reverse process. To further improve efficiency, the deterministic discrete integration method (DDIM) is employed~\cite{song2020denoising}, significantly reducing the sampling steps. Our model not only achieves state-of-the-art performance on two public datasets, but also surpasses other baselines in terms of efficiency and robustness of SNR. Overall, this algorithm marks a advancement in speech enhancement, optimizing performance and efficiency without the need for extra parameters or escalating computational complexity.

\section{Related work}

\subsection{Speech enhancement}
Traditional speech enhancement methods leverage the statistical features of the target and interference signals across various domains, such as time, spectrum, cepstrum and spatial distribution~\cite{gerkmann2018book_chapter}. Conversely, machine learning-based techniques aim to learn these statistical properties~\cite{wang2018supervised}. They can be categorized into discriminative and generative models. The former is dominated by predictive methods that use supervised learning to learn a single best deterministic mapping between corrupted speech $y$ and the corresponding clean speech target $x$~\cite{wang2018supervised}. 
Common approaches include tf masking~\cite{williamson2015complex}, complex spectral mapping~\cite{fu2017complex}, or operating directly in the time domain~\cite{fu2017raw}. Generative models, on the other hand, implicitly or explicitly learn the target distribution and allow to generate multiple valid estimates instead of a singular and optimal prediction as for predictive approaches~\cite{murphy2023probabilistic}. Generative approaches include VAE-based explicit density estimations~\cite{fang2021variational}, normalizing flows adding invertible transforms to obtain tractable marginal likelihoods~\cite{kobyzev2020normalizing}, GAN estimating implicit distributions~\cite{ kong2020hifi}, flow-based models~\cite{nugraha2020flow} and diffusion model-based approaches~\cite{tai2024dose,tai2023revisiting}. 

\subsection{Conditional diffusion models}
Diffusion models are adept at not only generating data from a distribution $p_0$ but can also from a specific conditional distribution $p_0(x|c)$ when provided with a condition $c$. These conditions can guide the generation process using specified conditional information, ensuring the data produced meets particular conditions or characteristics~\cite{yang2024survey}. Those guidances could be class labels or features related to the input data $x$. Additionally, there are specific sampling algorithms designed for conditional generation, such as label-based conditions~\cite{dhariwal2021diffusion}, self guidance~\cite{epstein2024diffusionself}, physical based guidance~\cite{yuan2023physdiff}, feature-based guidance~\cite{mei2024latent}, etc. These conditional mechanisms are more conducive to the generation of application-specific fields by using the control of other information to generate results. Currently, conditional diffusion model is also the most common method in various application scenarios due to its highly specific and controlled outputs.

\subsection{Denoising diffusion implicit model (DDIM)}
Denoising diffusion implicit models (DDIMs) are a variant of denoising diffusion probabilistic models (DDPMs)~\cite{song2020denoising}. DDIMs produce high-quality samples by gradually denoising data over a series of steps. Unlike traditional DDPMs that typically require a large number of steps to generate samples, DDIMs enable faster sampling with fewer steps without significantly compromising the quality of the generated data. In detail, they achieve this by modifying the diffusion process to be non-Markovian, allowing for an implicit inference of the denoising path, which in turn allows for more direct and efficient sampling paths from the noise distribution back to the data distribution. This modification leads to more efficient and practical applications in various fields such as image synthesis, text-to-image generation, and audio synthesis. By reducing the sampling time while maintaining output fidelity, DDIMs represent a significant advancement in the field of generative models, making them suitable for real-time applications and lowering computational costs.

\section{Proposed method}

\begin{figure*}[!t]
\centering
\includegraphics[width=0.96\textwidth]{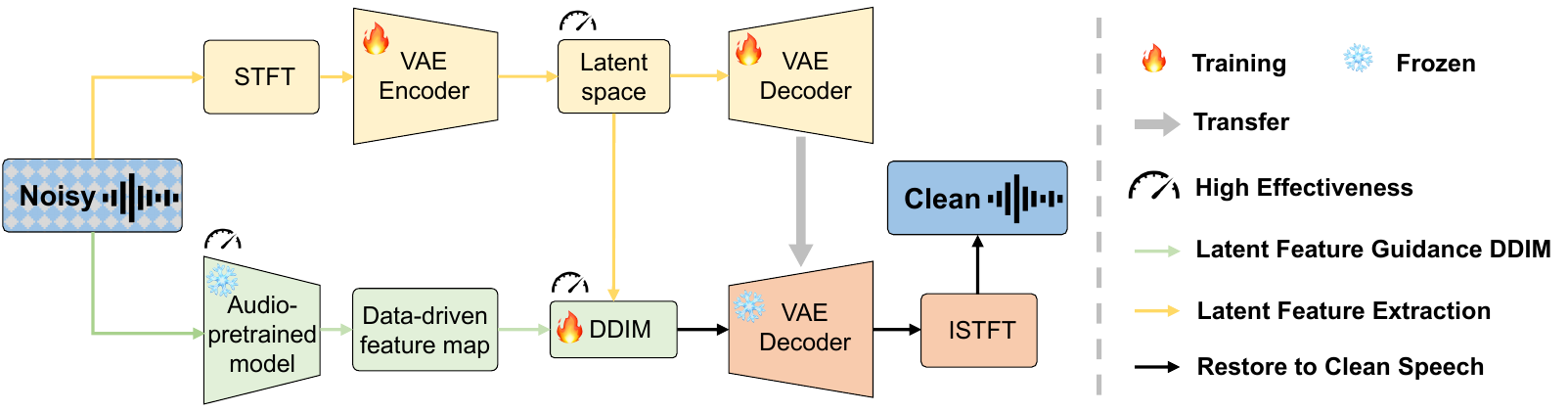}
\caption{The workflow of the proposed FUSE method. It consists of three processes: \textcolor{Yellow}{\textbf{latent feature extraction based on VAE}} (Sec.~\ref{sec3.1}), \textcolor{LimeGreen}{\textbf{pre-training feature guided diffusion model}} (Sec.~\ref{sec3.2}), and \textcolor{YellowOrange}{\textbf{restore to clean speech}} (Sec.~\ref{sec3.3}). Within these, we specifically highlight three components that can enhance the efficiency and the parts that need to be trained or frozen during training process.} 
\label{fig:workflow} 
\end{figure*}

In this section, we will introduce the proposed model, as illustrated in Fig.~\ref{fig:workflow}. It comprises three main components: 1) latent feature extraction based on VAE, 2) pre-training feature guided diffusion model, and 3) restore to clean speech.

\subsection{Latent feature extraction based on VAE} \label{sec3.1}
We first obtain the spectrogram through a sliding window and STFT, and input it to pass through the encoder of the VAE, mapping the spectrogram to a latent space corresponding to the variational distribution parameters. Then, we take this representation from the latent space as input for the following diffusion model. By reducing the shape, we enhance the efficiency of training within the diffusion model.
 
In detail, we treat the spectrogram obtained from STFT as an image $x$. Assume that the VAE encoder defines a conditional probability distribution $q_{\phi}(z|x)$, aimed at approximating the true posterior distribution $p(z|x)$. In the VAE, it is typically assumed to be Gaussian distribution, with its mean and variance parameterized by a neural network (the encoder network):
\begin{equation}
    \mathsf{Mean}: \mu = \mu_{\phi}(x); \quad \mathsf{Variance}: \sigma^2 = \sigma^2_{\phi}(x),
\end{equation}

\noindent where $\mu_{\phi}(x)$ and $\sigma^2_{\phi}(x)$ are the output of the encoder, and $\phi$ is the parameters the model need to learn. Then, for the backpropagation training, the reparameterization trick is applied to sample the latent variable $z$. It includes extracting the noise $\epsilon \sim \mathcal N(0,I)$ and transforming it using the following equation:
\begin{equation}
    z = \mu_{\phi}(x) + \sigma^2_{\phi}(x) \cdot \epsilon.
\end{equation}

As for the corresponding decoder, the training process expects decoder to learn a probability distribution $p_{\theta}(x|z)$, and $\theta$ is the learnable parameter of the decoder. For the loss function, we use the reconstruction-based loss (MSE) together with the KL divergence for training, which is often referred to as Evidence Lower Bound (ELBO):
\begin{equation}
    \mathsf{Loss} :=\mathsf{-ELBO} = -\mathbb{E}_{q_{\phi}(z|x)} [\mathsf{log}p_{\theta}(x|z)] + \mathsf{KL}[q_{\phi}(z|x)||p_{\theta}(z)].
\end{equation}

By gradient descent method, the network parameters of encoder $\phi$ and decoder $\theta$ are adjusted and the loss function is minimized to obtain the latent space representation $z$, which is used as the input of following diffusion model. In particular, the dimension of $z$ is much smaller than the dimension of the input $x$, which also ensures computational efficiency and optimizes complexity during the diffusion model training process.

\begin{algorithm}[!t]
\caption{Conditional DDIM Training Process}\label{alg:ddim_training}
\begin{algorithmic}[1]
\State Initialize the parameters of the model $\theta^{'}$
\State \textbf{Input:} VAE latent features $z$, condition $c$, noise levels $\{\alpha_1, \alpha_2, \ldots, \alpha_T\}$
\For{each training iteration}
    \State Sample a minibatch of latent features $\{z_i\}$ from $z$
    \State Sample a minibatch of conditions $\{c_i\}$ from $c$
    \State Sample noise level $\alpha_t$ uniformly at random
    \State Add noise to the data: $z^*_i = z_i + \alpha_t \cdot \epsilon_i, \epsilon_i \sim \mathcal{N}(0, I)$
    \State Compute loss: $\mathcal{L}(\theta^{'}) = \mathbb{E}\left[||f_{\theta^{'}}(z^*_i, c_i, \alpha_t) - \epsilon_i||^2\right]$
    \State Update the parameters $\theta^{'}$ to minimize $\mathcal{L}(\theta^{'})$ using gradient descent
\EndFor
\State \textbf{return} Trained model parameters $\theta^{'}$
\end{algorithmic}
\end{algorithm}

\begin{algorithm}[!t]
\caption{Conditional DDIM Sampling Process}\label{alg:ddim_sampling}
\begin{algorithmic}[1]
\State \textbf{Input:} Trained model parameters $\theta^{'}$, condition $c$, initial noise level $\alpha_T$
\State Initialize $z^*_T$ with noise: $z^*_T\sim \mathcal{N}(0, I)$
\For{$t = T$ \textbf{to} $1$}
    \State Compute the noise level $\alpha_{t-1}$ for the previous step
    \State Estimate the original noise $\hat{\epsilon}_t = f_{\theta^{'}}(z^*_t, c, \alpha_t)$
    \State Compute the denoised sample for the next step:
    \State $z^{*}_{t-1} = \frac{z^{*}_t - \alpha_t \cdot \hat{\epsilon}_t}{\sqrt{1 - \alpha_t^2}}$
\EndFor
\State \textbf{return} The generated data $z^*$
\end{algorithmic}
\end{algorithm}

\subsection{Pre-training feature guided diffusion model} \label{sec3.2}
We first employ an audio-pretrained model to extract general acoustic features, and they serve as conditions for the following diffusion model. This approach has two main purposes: 1) Enhance the efficiency of feature extraction. We do not need to fine-tune the pre-trained model, so we directly use a frozen encoder to obtain the acoustic representations. 2) Since the model is trained with a more diversified data, its representations can provide guidance to the diffusion model for generation in the correct direction, differing from the VAE's latent space features. 

Specifically, we use BEATs~\cite{chen2022beats} as the backbone of pre-training model. Extract data-driven representations from segmented audio $x_{ori}$, denoted as $c = \mathsf{BEATs}(x_{ori})$. Due to its use of a complete pre-module, it requires no additional weights, making it highly efficient. Then, this data-driven feature $c$ will be used as condition of the diffusion model. 

As for DDIM, the forward process, which does not depend directly on conditional guidance, can be defined as,
\begin{equation} \label{DDIM1}
    q(z_t|z_0) =  \mathcal N(z_t; \sqrt{ {\alpha}_{t}}x_0,(1- {\alpha}_{t})I),
\end{equation}

\noindent where the $ {\alpha}_t \in (0,1)$ decreases as $t$ increases. The joint distribution of DDIM is:
\begin{equation}
    q(z_{1:T}|z_0) =  q(z_T|z_0) \prod\limits_{t=2}^T q(z_{t-1}|z_t,z_0).
\end{equation}

\noindent Combining this with Eq.~\ref{DDIM1}, $z_t$ at any step can be obtained.

The reverse process can be written as ordinary differential equation (ODE) without the limitation of Markov assumption:
\begin{align}
    z_{t}=&\frac{\sqrt{\alpha_{t}}}{{\sqrt{\alpha_{t-1}}}}z_{t-1} + \label{eq:ddim_forward_exact}\\
    &\sqrt{\alpha_{t}} \left(\sqrt{\frac{1}{\alpha_{t}}-1} - \sqrt{\frac{1}{\alpha_{t-1}}-1}\right)\epsilon_{\theta^{'}}(z_{t}, t, c),\nonumber
\end{align}
\noindent where $c$ is the condition from pretrained model feature map. $z_t$ appears at both sides, so we estimate $\epsilon_{\theta^{'}}(z_{t}, t, c)$ with $\epsilon_{\theta^{'}}(z_{t-1}, t-1, c)$ by Euler method. Finally, compute the genertive latent space features $z^*$ at each step, that is:
\begin{align}
    z_{t}^*=&\frac{\sqrt{\alpha_{t}}}{{\sqrt{\alpha_{t-1}}}}z_{t-1}^* + \label{eq:ddim_forward}\\
    &\sqrt{\alpha_{t}} \left(\sqrt{\frac{1}{\alpha_{t}}-1} - \sqrt{\frac{1}{\alpha_{t-1}}-1}\right)\epsilon_{\theta^{'}}(z_{t-1}^*, t-1, c).\nonumber
\end{align}

The training and sampling process of the proposed conditional DDIM are shown in Alg.~\ref{alg:ddim_training} and Alg.~\ref{alg:ddim_sampling}.

\subsection{Restore to clean speech} \label{sec3.3}
In the final process, we decode the generated latent feature $z^{*}$ by a previously learned frozen decoder and convert this to clean speech by inverse short time Fourier transform (ISTFT).

\begin{table*}[!t]
    \centering
    \caption{Speech enhancement results of two datasets average of five times. Values indicate mean and standard deviation, respectively. Methods are sorted by the algorithm type, generative (G) or discriminative (D). In the \sethlcolor{lightgray} \hl{gray background} is our proposed method.}
\centering
\resizebox{0.95\textwidth}{!}{
\begin{tabular}{l|cc|ccccccc}
\toprule
\textbf{Method} & \textbf{Type} & \textbf{Training set} & \textbf{POLQA $\uparrow$} & \textbf{PESQ $\uparrow$} & \textbf{ESTOI $\uparrow$} & \textbf{SI-SDR [dB] $\uparrow$} & \textbf{SI-SIR [dB] $\uparrow$} & \textbf{SI-SAR[dB] $\uparrow$} & \textbf{DNSMOS $\uparrow$} \\
\midrule
STCN \cite{richter2020speech}& G & WSJ0 & $2.64 \pm 0.68$ & $2.01 \pm 0.55$ & $0.81 \pm 0.12$ & $13.5 \pm 4.7$ & $18.7 \pm 5.5$ & $15.4 \pm 4.7$ & $3.34 \pm 0.37$ \\
CDiffuse \cite{lu2022conditional}& G & WSJ0-C3 & $3.08 \pm 0.58$ & $2.27 \pm 0.51$ & $0.83 \pm 0.09$ & $\ \ 9.2 \pm 2.3$ & $19.8 \pm 5.9$ & $10.0 \pm 2.3$ & $3.43 \pm 0.32$ \\
DVAE \cite{richter2020speech}& G & WSJ0 & $2.97 \pm 0.63$ & $2.31 \pm 0.55$ & $0.85 \pm 0.11$ & $15.8 \pm 5.0$ & $21.6 \pm 6.1$ & $17.6 \pm 4.9$ & $3.61 \pm 0.29$ \\
SGMSE \cite{welker2022speech}& G & WSJ0-C3 & $2.98 \pm 0.60$ & $2.28 \pm 0.57$ & $0.86 \pm 0.09$ & $14.8 \pm 4.3$ & $25.4 \pm 5.6$ & $15.3 \pm 4.2$ & $3.70 \pm 0.27$ \\
SGMSE+ \cite{welker2022speech}& G & WSJ0-C3 & $3.73 \pm 0.53$ & $2.96 \pm 0.55$ & $0.92 \pm 0.06$ & $18.3 \pm 4.4$ & $31.1 \pm 4.6$ & $18.6 \pm 4.5$ & ${3.99 \pm 0.19}$ \\

\rowcolor{lightgray} \textbf{FUSE (Step=6)} & G & WSJ0-C3 & $\mathbf{3.91 \pm 0.55}$ & $\mathbf{3.13 \pm 0.54}$ & ${0.91 \pm 0.07}$ & ${19.7 \pm 4.2}$ & $\mathbf{31.5 \pm 5.7}$ & $\mathbf{21.2 \pm 4.8}$ & $\mathbf{4.10 \pm 0.25}$ \\

\midrule
MetricGAN+ \cite{fu2021metricgan+}& D & WSJ0-C3 & $3.52 \pm 0.61$ & ${3.03 \pm 0.45}$ & $0.88 \pm 0.08$ & $10.5 \pm 4.5$ & $24.5 \pm 5.1$ & $10.7 \pm 4.6$ & $3.67 \pm 0.30$ \\
Conv-TasNet \cite{luo2019conv}& D & WSJ0-C3 & $3.65 \pm 0.54$ & $2.99 \pm 0.58$ & $\mathbf{0.93 \pm 0.05}$ & $\mathbf{19.9 \pm 4.3}$ & $29.2 \pm 4.6$ & ${20.6 \pm 4.5}$ & $3.79 \pm 0.27$ \\
\midrule
\midrule

STCN \cite{richter2020speech}& G & VB & $2.53 \pm 0.66$ & $1.80 \pm 0.45$ & $0.79 \pm 0.12$ & $11.9 \pm 4.5$ & $17.3 \pm 4.9$ & $13.8 \pm 4.6$ & $3.40 \pm 0.34$ \\
CDiffuse \cite{lu2022conditional}& G & VB-DMD & $2.15 \pm 0.57$ & $1.79 \pm 0.42$ & $0.71 \pm 0.11$ & $ \ \ 3.2 \pm 3.2$ & $21.8 \pm 7.0$ & $\ \ 3.4 \pm 3.2$ & $3.17 \pm 0.29$ \\
DVAE \cite{richter2020speech}& G & VB & $2.84 \pm 0.61$ & $2.08 \pm 0.49$ & $0.82 \pm 0.11$ & $13.9 \pm 4.8$ & $19.5 \pm 5.9$ & $15.8 \pm 4.7$ & $3.52 \pm 0.31$ \\
SGMSE \cite{welker2022speech}& G & VB-DMD & $2.66 \pm 0.58$ & $1.94 \pm 0.47$ & $0.81 \pm 0.11$ & $13.3 \pm 4.3$ & $23.5 \pm 6.0$ & $13.8 \pm 4.2$ & $3.76 \pm 0.25$ \\
SGMSE+ \cite{welker2022speech}& G & VB-DMD & ${3.43 \pm 0.61}$ & ${2.48 \pm 0.58}$ & $\mathbf{0.90 \pm 0.07}$ & ${16.2 \pm 4.1}$ & $\mathbf{28.9 \pm 4.6}$ & ${16.4 \pm 4.1}$ & ${4.00 \pm 0.19}$ \\

\rowcolor{lightgray} \textbf{FUSE (Step=6)} & G & VB-DMD & $\mathbf{3.80 \pm 0.64}$ & $\mathbf{2.78 \pm 0.55}$ & ${0.86 \pm 0.10}$ & $\mathbf{17.0 \pm 3.6}$ & $\mathbf{28.9 \pm 6.6}$ & $\mathbf{16.8 \pm 4.1}$ & $\mathbf{4.18 \pm 0.18}$ \\

\midrule
MetricGAN+ \cite{fu2021metricgan+}& D & VB-DMD & $2.47 \pm 0.67$ & $2.13 \pm 0.53$ & $0.76 \pm 0.12$ & $\ \ 6.8 \pm 3.1$ & $22.9 \pm 4.9$ & $\ \ 7.0 \pm 3.1$ & $3.51 \pm 0.29$ \\
Conv-TasNet \cite{luo2019conv}& D & VB-DMD & $3.13 \pm 0.60$ & $2.40 \pm 0.53$ & $0.88 \pm 0.08$ & $15.2 \pm 3.9$ & $26.5 \pm 4.6$ & $15.6 \pm 4.0$ & $3.68 \pm 0.30$ \\
\bottomrule
\end{tabular}} 
\vspace{-1em}
\label{tab:result1_diff4SE}
\end{table*}

\section{Experiment}

\subsection{Experimental setup}

\textbf{Datasets}: We use WSJ0-CHiME3 and VoiceBank-DEMAND datasets for the evaluation. We employ two datasets facilitates cross-dataset evaluation, wherein testing is carried out on a different dataset than the one utilized for training. 

\textbf{Evaluation metrics}: We use seven different standard metrics for fair assessment, including full reference algorithms (POLQA, PESQ, ESTOI, SI-SDR, SI-SIR, and SI-SAR) and non-intrusive metrics (DNSMOS). 

\textbf{Compared baselines}: We compared our proposed method with other five generative methods (STCN~\cite{richter2020speech}, CDiffuse~\cite{lu2022conditional}, DVAE~\cite{bie2022unsupervised}, SGMSE, and SGMSE+~\cite{welker2022speech}) and two discriminative methods (MetricGAN+~\cite{fu2021metricgan+} and Conv-TasNet~\cite{luo2019conv}).

\textbf{Hyperparameters}: The sampling rate of raw audio is 16 kHz. As for the STFT, the hop length between consecutive windows is set to 128 samples, and Hann window is used for smoothing. We standardize the spectrograms by trimming them to $T=256$ time frames. We set the step of reverse process to 6 to make the algorithm as efficient as possible.

\textbf{Deployment}: We train proposed model on four NVIDIA A10 (24GB). Adam optimizer is utilized. The initial learning rate is 0.0005, with a decay to 95\% of every 10 epochs. A batch size of 32 and a training duration of 100 epochs were selected. The model was developed using PyTorch 1.12 under Python 3.9.

More experimental details are shown in the Appendix.

\begin{figure}[!t]
\centering
\includegraphics[width=0.9\columnwidth]{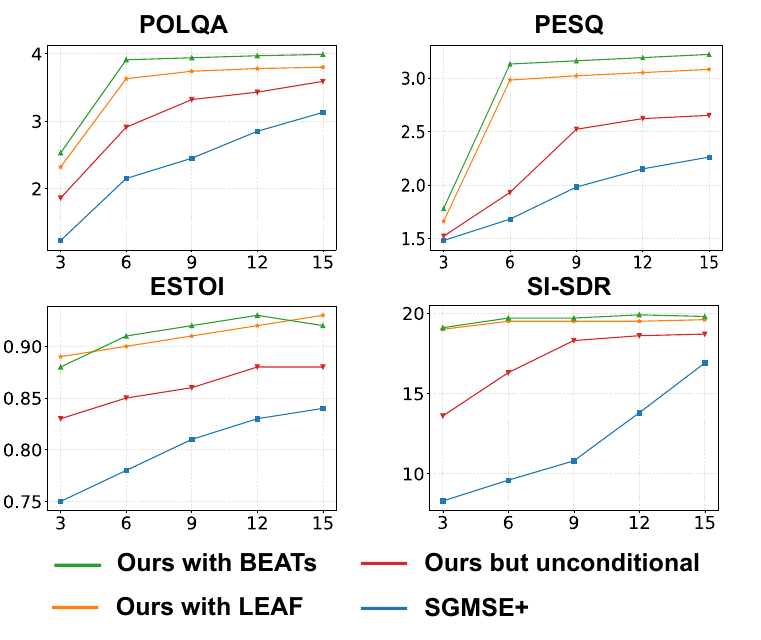} 
\caption{The results of different pretrained features and unconditional features training on WSJ0-CHiME3 dataset as a function of the number of reverse diffusion steps.} 
\label{fig:abtest}  \vspace{-1.2em}
\end{figure}

\subsection{Result}
The detailed results for WSJ0-CHiME3 and VoiceBank-DEMAND datasets are shown in Tab.~\ref{tab:result1_diff4SE}. Compared to all other methods, our proposed approach demonstrates superior performance across most metrics on two datasets, particularly in indices such as POLQA, PESQ, and DNSMOS. This indicates our method's significant effectiveness in enhancing speech clarity and overall quality. Notably, we only use six steps in the reverse process of the diffusion model, significantly improving efficiency compared to other methods. Moreover, our dataset features randomly varied SNR, meaning it includes different levels of noise interference, yet the variance of our method's results remains minimal, proving its robustness. In summary, our method is capable of providing high-quality and efficient speech enhancement in complex noise environments.

In addition, we did some ablation studies to test the effect of 1) w/wo conditional guidance, 2) different conditional guidances, and 3) reverse diffusion step on the results, which are shown in Fig.~\ref{fig:abtest}. We utilized the WSJ0-CHiME3 dataset for training. The results indicate that unconditional methods were not as effective and required more time to reach a stable state. In contrast, conditional guided methods needed only a few steps to generate speech of comparatively high quality. Additionally, LEAF~\cite{zeghidour2021leaf}, another pre-trained model, showed results that were not significantly different from those guided by BEATs. Besides, compared to other methods, our approach achieves better outcomes with fewer steps, demonstrating its efficiency.

\section{Conclusion}
This paper presented a novel pretraining feature-guided unified diffusion model, named \textbf{FUSE}, aimed at addressing the efficiency and performance challenges in speech enhancement. By integrating two complementary acoustic features and employing DDIM, our model significantly reduces sampling steps while maintaining high-quality speech synthesis. Our approach demonstrated state-of-the-art results on two public datasets. It outperforms existing baselines in both efficiency and robustness, offering a promising solution to the limitations of diffusion-based methods in real-world applications.

\clearpage
\bibliography{7_reference}
\bibliographystyle{IEEEtran}

\appendix
\section{Appendix}
\subsection{Datasets}
To assess the effectiveness of our speech enhancement strategy, we employ two distinct datasets: the WSJ0-CHiME3 and the VB-DMD. These datasets serve as the basis for our cross-dataset evaluation approach, where models trained on one dataset are tested on the other. This methodology allows us to examine the adaptability of our enhancement method to unfamiliar data exhibiting varied noise types and recording conditions, thus providing insights into the generalization capabilities of the proposed approach. This cross-dataset testing framework is particularly valuable for understanding the robustness of our method across diverse acoustic environments, further illuminating its potential for real-world application.

\textbf{WSJ0-CHiME3} The WSJ0-CHiME3 dataset is a hybrid collection that merges clean speech utterances from the Wall Street Journal (WSJ0) corpus \cite{datasetWSJ0} with environmental noise samples from the CHiME3 dataset \cite{barker2015third}. To create a mixed signal, a noise file is randomly chosen and overlaid onto a clean speech utterance, ensuring each utterance is utilized uniquely. The Signal-to-Noise Ratio (SNR) for the assembled training, validation, and testing sets is uniformly distributed, ranging from 0 to 20 dB. This dataset provides a realistic and challenging environment for evaluating speech enhancement algorithms, facilitating the assessment of their robustness and effectiveness in varied acoustic conditions.

\textbf{VB-DMD}: For our research, we employed the widely recognized VoiceBank-DEMAND dataset (VB-DMD)~\cite{valentini2016investigating}, a standard benchmark for evaluating single-channel speech enhancement techniques. This dataset comprises utterances that have been artificially mixed with a variety of noise types to simulate real-world conditions. Specifically, eight genuine noise recordings from the DEMAND dataset~\cite{thiemann2013diverse} and two synthetic noise types were used to contaminate the speech samples. These mixtures were created at four different Signal-to-Noise Ratios (SNRs): 0, 5, 10, and 15 dB to train the models under various degrees of noise interference.

\subsection{Evaluation Metrics}
To assess the effectiveness of our proposed approach, we apply a suite of standard metrics, elaborated upon further below. Metrics (a) through (d) utilize full-reference algorithms, quantitatively comparing the processed signal against a clean reference signal through established digital signal processing techniques. Conversely, metric (e) employs a non-intrusive evaluation method, suitable for analyzing real-world recordings in scenarios where a clean reference signal is not accessible. This comprehensive evaluation framework allows for a nuanced understanding of the method's performance across a variety of conditions, highlighting its applicability and robustness in practical settings.

\textbf{Perceptual Objective Listening Quality Analysis (POLQA)}: POLQA, a standard established by ITU-T, utilizes a perceptual model to estimate speech quality \cite{polqa2018}. Its scoring system ranges from 1, indicating poor quality, to 5, signifying excellent quality, offering a comprehensive measure for speech assessment.

\textbf{Perceptual Evaluation of Speech Quality (PESQ)}: Serving as an objective benchmark for speech quality analysis, PESQ is recognized under ITU-T P.862 \cite{rixPerceptualEvaluationSpeech2001}. Despite POLQA succeeding it, PESQ remains prevalent in academic circles. Scores span from 1 (poor) to 4.5 (excellent), with distinctions made between wideband and narrowband assessments.

\textbf{Extended Short-Time Objective Intelligibility (ESTOI)}: ESTOI offers a quantitative measure for speech intelligibility when faced with various distortions \cite{jensen2016algorithm}. Its normalized scoring from 0 to 1 helps evaluate the clarity of speech, where higher scores indicate greater intelligibility.

\textbf{Scale-Invariant Metrics (SI-SDR, SI-SIR, SI-SAR)}: These metrics, crucial for evaluating speech enhancement and separation in single-channel settings, include the Signal-to-Distortion Ratio (SDR), Signal-to-Interference Ratio (SIR), and Signal-to-Artifact Ratio (SAR) \cite{leroux2018sdr}. Expressed in dB, higher scores reflect superior audio quality.

\textbf{Deep Noise Suppression Mean Opinion Score (DNSMOS)}: A novel, reference-free evaluation metric, DNSMOS leverages a deep neural network trained on human listening ratings to gauge speech quality \cite{reddy2021dnsmos}. It draws on expansive listening experiments conducted online, adhering to the ITU-T P.808 standards \cite{itu-t-rec808,naderi2020open}, providing a modern approach to speech quality assessment.

These metrics collectively offer a robust framework for the comprehensive evaluation of speech quality, intelligibility, and the effectiveness of enhancement algorithms, facilitating a deeper understanding of their impact across a variety of scenarios.

\subsection{Baselines}
In our analysis, we benchmark the effectiveness of our newly introduced method against a mix of six baselines, comprising both generative and discriminative models, elaborated upon further below.

\textbf{STCN} \cite{richter2020speech}: This generative method for speech enhancement leverages a Variational Autoencoder (VAE) framework, incorporating a Stochastic Temporal Convolutional Network (STCN) to manage the latent variables with hierarchical and temporal dependencies, optimized through a Monte Carlo expectation-maximization algorithm.

\textbf{DVAE} \cite{bie2022unsupervised}: An innovative generative technique utilizing an unsupervised Dynamical VAE (DVAE) to capture the temporal links between successive observable and latent variables, with parameters refined during testing via a variational expectation maximization method, including encoder adjustments through stochastic gradient ascent.

\textbf{CDiffuSE} \cite{lu2022conditional}: CDiffuSE is a generative model enhancing speech through a conditional diffusion process in the time domain, showcasing the adaptability and effectiveness of diffusion processes in handling complex auditory data.

\textbf{SGMSE} \cite{welker2022speech}: The Score-based Generative Model for Speech Enhancement (SGMSE) employ a deep complex U-Net for the score model.

\textbf{MetricGAN+} \cite{fu2021metricgan+}: A leading discriminative model that optimizes a generative network for mask-based prediction of clean speech, paired with a discriminator trained to estimate the PESQ score, bridging the gap between prediction accuracy and perceptual quality.

\textbf{Conv-TasNet} \cite{luo2019conv}: An end-to-end architecture that computes a mask for filtering through a learned representation of noisy mixtures, with the refined signal then reconverted to the time domain using a custom decoder, illustrating the power of end-to-end learning in audio processing.

Through comprehensive retraining and analysis, our goal was to rigorously compare these methodologies, across a spectrum of conditions to delineate the nuances of each approach in enhancing speech quality effectively.

\end{document}